\documentclass{iau}
\usepackage{multirow}
\usepackage{multicol}

\begin{document}

\lefttitle{Yeimy J. Rivera}
\righttitle{Sun to heliosphere integrated science}

\jnlPage{1}{7}
\jnlDoiYr{2024}
\doival{10.1017/xxxxx}
\volno{390}
\pubYr{2024}
\journaltitle{Multi-Point view of the Sun: Advances in Solar Observations and in Space Weather Understanding}

\aopheadtitle{Proceedings of the IAU Symposium}
\editors{M. Romoli, L. Feng, M. Snow, eds.}

\title{An assessment of observational coverage and gaps for robust Sun to heliosphere integrated science}

\newcommand{\CfA}{\affiliation{Center for Astrophysics $|$ Harvard \& Smithsonian\\
		60 Garden Street, Cambridge, MA 02138, USA}}

\NewDocumentCommand{\PlaceAffil}{m}{\textsuperscript{#1}}
\author{Yeimy~J.~Rivera\PlaceAffil{1}, Samuel T. Badman\PlaceAffil{1}}
\CfA

\begin{abstract}

Understanding the generation and development of the continuous outflow from the Sun requires tracing the physical conditions from deep in the corona to the heliosphere. Detailed global observations of plasma state variables and the magnetic field are needed to provide critical constraints to the underlying physics driving models of the corona and solar wind. Key diagnostics of the solar wind require measurements at its formation site and during its outflow to continuously track it across rapidly changing regions of space. A unified view of the solar wind is only possible through coordinated remote and in situ observations that probe these different regions. Here, we discuss current observational coverage and gaps of different plasma properties and review recent coordinated studies. We highlight how these efforts may become more routine with the launch of upcoming and planned missions.

\end{abstract}

\maketitle

\section{Introduction \label{sec:intro}}

The solar wind is a continuous flow of plasma that fills the heliosphere. However, the connection between its formation, development, and local, ongoing plasma dynamics as it streams from the Sun is known partially or in segments. A full theory of the solar wind requires extended coronal remote observations that track its birth in the corona and as it escapes into the inner heliopshere. 

Recent observations and modeling of solar wind formation have shown that previously unresolved reconnection events in polar coronal holes are likely driving the formation of the fast solar wind \citep{Chitta2023Sci, Bale2023, Raouafi2023}. Ubiquitous brightenings in the extreme ultraviolet (EUV) within the inter-plume regions of coronal holes suggest that small-scale reconnection is taking place low in the corona in seemingly open-field structures. Continuous interchange reconnection is thought to be the main contribution to the fast solar wind however direct mapping of the solar wind to the coronal dynamics has not been well established. On the other hand, the slow solar wind coronal sources are thought to be more diverse and, likely, contain changing contributions from different sources throughout the solar cycle \citep{Abbo2016}. Recently, extended coronal observations in the EUV indicate that slow wind formation and outflow is connected to S-web arcs \citep{Antiochos2011_sweb, Chitta2023NatAs} across the middle corona \citep{West2023}. Strong outflows at the periphery of active regions (ARs) have also been identified as important drivers of slow speed streams \citep{Brooks2020}. All processes suggested likely provide some contribution to the slow solar wind as well as to its variability measured in situ.

A key to bridging coronal and heliospheric physics lies in our ability to continue tracing the newly formed solar wind stream beyond the solar atmosphere. After leaving the highly structured corona, a broad spectrum of properties of the solar wind continue to evolve in different ways as observed in situ across the heliosphere. From decades of measurements of solar wind taken from near 1 AU, the plasma state of the fastest speed solar wind is found to be generally less variable in various properties compared to slower speed wind, e.g. temperature, density, Alfv\'enic character, elemental and ion composition, heavy ion differential flow, entropy, while the slower solar wind exhibits properties spanning a much larger range, often with significant overlap with the properties of fast wind streams \citep[see e.g. Figure 1 in][]{Rivera2025}. There is also strong evidence that solar wind properties non-uniformly change with distance from the Sun as the plasma expands, which makes identifying and connecting solar wind at different stages of its evolution very complex \citep{Marsch1982, Bruno2013}. For instance, several studies have revealed that the solar wind continues to be heated and accelerated well outside the corona, as indicated by its non-adiabatic temperature profile \citep{Richardson2003, Hellinger2011}. The polytropic index that describes the temperature behavior of the solar wind can vary with speed and between protons and electrons \citep{Dakeyo2022}. Recent studies reveal that the drivers of solar wind acceleration can be distinct for different speed wind, where the fastest wind streams indicate a large contribution to its extended heating and acceleration, beyond the Alfv\'en surface, from Alfv\'enic fluctuations and their associated pressure gradient \citep{Halekas2023,Rivera2024}. At the same range of distances, the slower speed wind's acceleration can be fully accounted for through the non-adiabatic thermal pressure gradient of the proton and electrons collectively \citep{Halekas2022, Rivera2025}. However, important questions remain; how is the non-adiabatic temperature profile and additional heating observed in the ions and electrons produced between the corona and heliosphere? What is the role that Alfv\'en waves/turbulence play in the heating process? What are the conditions of their solar sources that drive their distinct evolution? Are the physical heating mechanisms in the solar wind distinct from those that produce coronal heating?

The coronal environment, formation, and continued evolution of the solar wind are intimately connected \citep{Viall2020JGRA_nine}. Traditionally, outflowing solar wind measured in situ can be mapped to their coronal sources using simple ballistic projections along the Parker Spiral to the source surface of the Sun ($\sim2.5R_{\odot}$), where a Potential-field Source-surface (PFSS) model of the extrapolated corona can be used to map the flow's photospheric footpoints \citep{Badman2020}. Similarly, in more sophisticated Magnetohydrodynamic (MHD) models, solar wind can be traced along flux tubes through a coupled heliospheric and coronal 3D domain to a coronal source \citep{Riley2019_PSP_predict, vanderHolst2019_PSP_predict}. While the MHD solutions allow for a more detailed comparison between the modeled and simulated conditions that inform on the underlying physics and model performance, both techniques have been shown to yield similar solar wind footpoint predictions \citep{Riley2006_PFSS_MHD, Badman2023JGRA, Ervin2024}. However, identifying clear transient solar wind creation in the corona, such as reconnection-driven outflows (as in the EUV brightening generating jetlets and picoflares), remains difficult to map to specific instances in the solar wind measured in situ. A clear incompatibility in connection studies lies in that the corona is inherently dynamic at many scales, while techniques for tracing outflows to/from the corona rely on a steady-state coronal configuration, making it difficult to study solar wind formation. The recent time-dependent MHD simulations showcased during the 2024 total solar eclipse predictions{\footnote{\url{https://www.predsci.com/corona/apr2024eclipse/home.php}}} from Predictive Science Incorporated demonstrate great promise towards linking the energization of the corona to the processes that heat the corona and drive the solar wind \citep{Linker2024EGU_eclipse}.

Historically, there has been a large gap between coronal observations and in situ heliospheric observations simply by virtue of the large spatial separation of these measurements. With the launch of Parker Solar Probe \citep{Fox2016}, the heliosphysics community now has in situ access to heliocentric distances just beyond where the solar wind is born and through the early stages of heating and acceleration experienced just beyond the corona. By surveying this region, several new points of constraint can be implemented over this regime of space with coordinated observations that collectively observe the formation and inner heliospheric evolution of the same solar wind stream. Similarly, Solar Orbiter \citep{Muller2020} will soon be adding new constraints over a wider range of heliographic latitude between the Earth and Sun. However, the utility of in situ measurements are limited when used in isolation due to their localization to single points in space. To go further in understanding the solar wind, ideal relative spacecraft configurations, appropriate modeling and remote observations of the solar wind must all be combined to link its creation and energization as it is imaged remotely and intercepted by orbiting spacecraft. 

\section{Solar Missions}
Our community, now more than ever is equipped to trace formation and energetics of out-flowing solar wind that permeates the solar system.  A continuous assessment of solar wind conditions and non-thermal features between its solar source and inner heliospheric evolution is critical for linking coronal to heliospheric dynamics and energy flow. Following the solar wind development from its point of origin will deepen our understanding of the sources, development, and dissipation of Alfv\'enic fluctuations, as well as the continuous role they play in heating and accelerating the corona and solar wind.

\begin{figure}
\begin{centering}
\includegraphics[width=\linewidth]{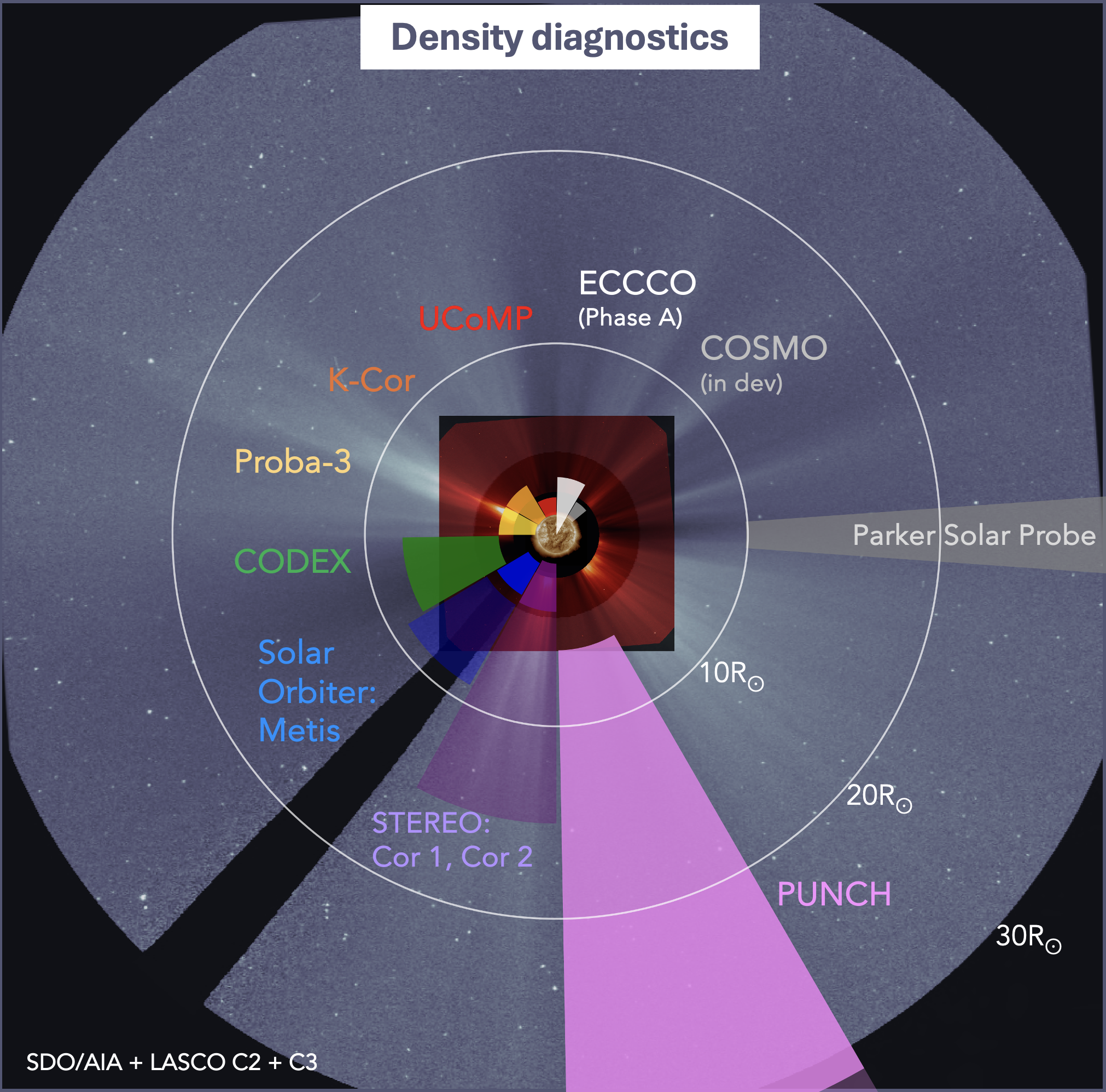}
\label{fig:FOVdensity_inner}
\caption{The image contains the field of view (FOV) of different ground and space-based instruments for density diagnostics of the low to outer corona described in Section \ref{sec:solarmissions}. Although we only use a small wedge to illustrate radial coverage, all instruments take remote observations of the full corona. We also include in situ coverage from Parker that ranges $\pm4^{\circ}$ from the ecliptic plane. The background image is a composite of SDO/AIA 193\AA~with LASCO C2 \& C3 encompassing a FOV of 30R$_{\odot}$. }
\end{centering}
\end{figure}

\begin{figure}
\begin{centering}
\includegraphics[width=\linewidth]{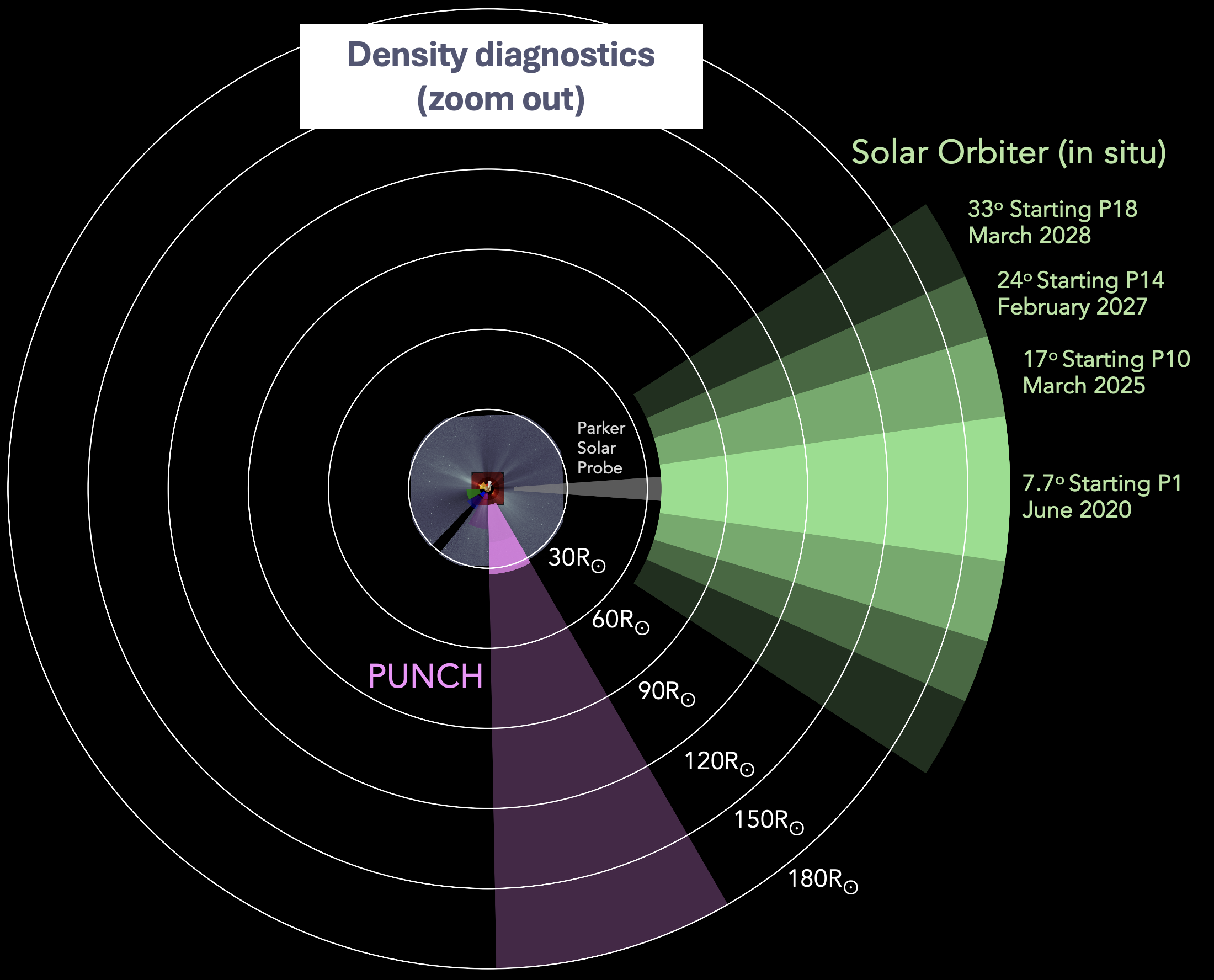}
\label{fig:FOVdensity_outer}
\caption{The image shows a zoomed out version of Figure \ref{fig:FOVdensity_inner} showing density diagnostics beyond 30R$_{\odot}$. The figure indicates the FOV remotely observed with PUNCH as well as the radial and latitudinal in situ range from Parker and Solar Orbiter. For Solar Orbiter, we include the growing latitudinal coverage starting after perihelion 10 (P10).}
\end{centering}
\end{figure}

\begin{figure}
\begin{centering}
\includegraphics[width=\linewidth]{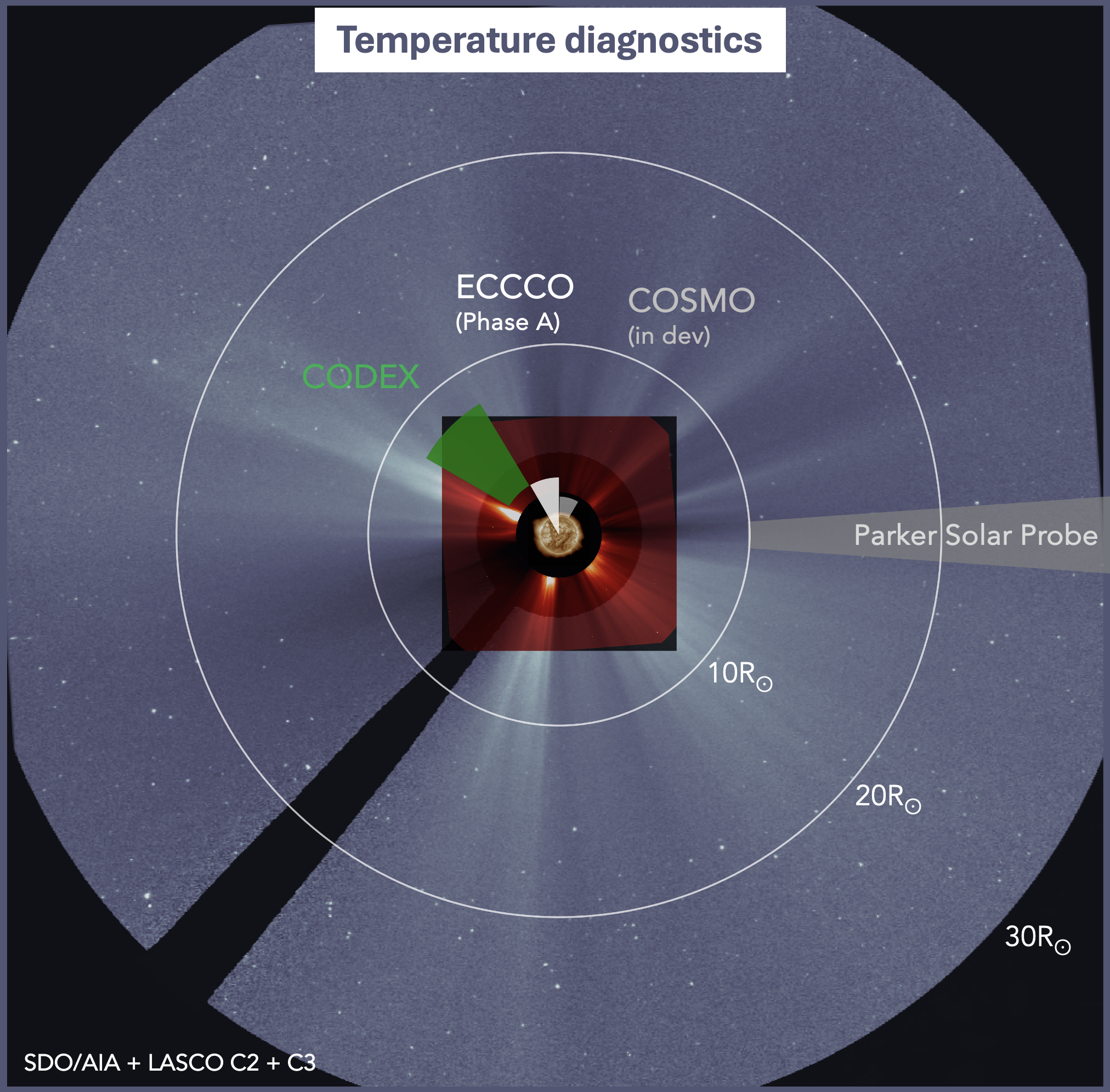}
\label{fig:FOVtemp_inner}
\caption{Same as Figure \ref{fig:FOVdensity_inner} for temperature (ion and electron) diagnostics of the low to outer corona. CODEX, ECCCO, and COSMO provide electron temperature diagnostics while Parker provides proton, alpha, and electron temperature.}
\end{centering}
\end{figure}

\begin{figure}
\begin{centering}
\includegraphics[width=\linewidth]{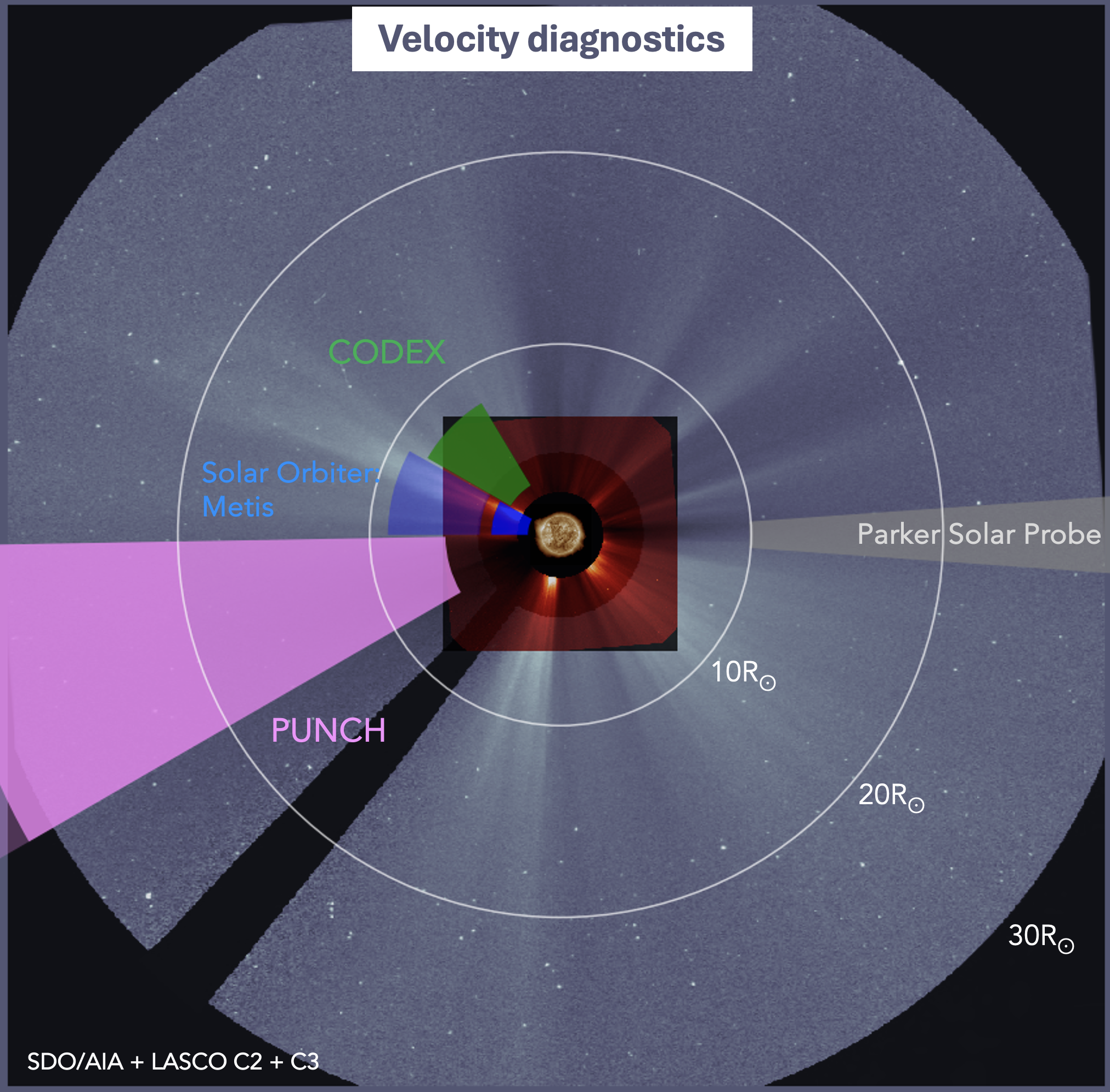}
\label{fig:FOVspeed_inner}
\caption{Same as Figure \ref{fig:FOVdensity_inner} for velocity diagnostics of the low to outer corona. CODEX provides electron velocity diagnostics, Metis and PUNCH provides proton velocity, and Parker measures proton and alpha velocities.}
\end{centering}
\end{figure}

\begin{figure}
\begin{centering}
\includegraphics[width=\linewidth]{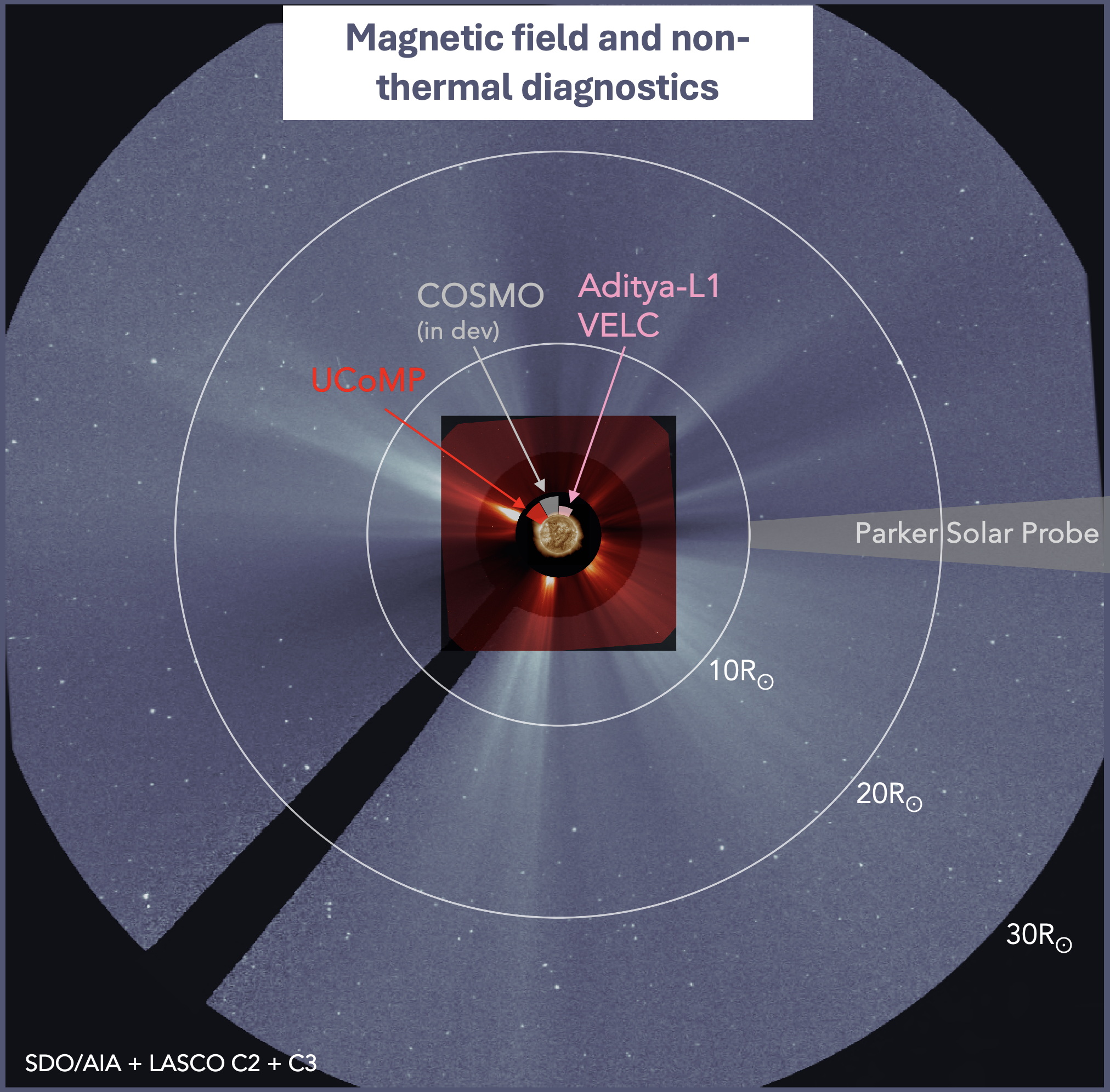}
\label{fig:FOVmagnetic_inner}
\caption{Same as Figure \ref{fig:FOVdensity_inner} for magnetic field and non-thermal diagnostics of the low to outer corona. }
\end{centering}
\end{figure}

This section briefly summarizes some diagnostics in the corona and solar wind that can be made and connected near-contemporaneously with current and planned missions. We restrict this section to include active and near-future missions with routine observations and full coverage of the corona (for remote sensing) that can capture the structure, formation, and outflow of the solar wind out to the closest in situ measurements. We include diagnostics of ambient conditions, and non-thermal characteristics that can result from wave, turbulence and instability processes important to constraining models of the solar wind.  Figure \ref{fig:FOVdensity_inner}, \ref{fig:FOVdensity_outer}, \ref{fig:FOVtemp_inner}, \ref{fig:FOVspeed_inner}, \ref{fig:FOVmagnetic_inner} summarize the field of view (FOV) of different instruments and spacecraft that can directly probe the (ion or electron) density, temperature, outflow speed, and magnetic field properties/non-thermal features in the corona and inner heliosphere. We note that, while remote observations provide global coronal diagnostics, the inversion of physical parameters from coronal emission is less direct than in situ measurements and usually present challenges from line of sight (LOS) integration effects. Conversely, in situ observations provide highly detailed and comprehensive diagnostics of the solar wind but only for a single point in space. Therefore, merging measurements can be non-trivial but highly valuable. We also note that non-thermal features in remote observations are typically expressed as non-thermal broadening and non-Maxwellian spectral line profiles observed in coronal emission while equivalent non-thermal characteristics in situ are exhibited by particle distributions deviating from Maxwellian profiles and direct measurements of fluctuations. The following sections briefly describe each of the diagnostics summarized in the figures from individual missions/observatories. 

\subsection{In situ Inner Heliospheric Observations} \label{sec:solarmissions}

There are two dedicated solar probes orbiting the Sun and surveying the solar wind below the distance of 1 au and off the Sun-Earth line: NASA's Parker Solar Probe (Parker) and ESA/NASA's Solar Orbiter. However, there have been studies that couple observations from solar and planetary probes in cruise phase to study solar phenomena at these distances as well, e.g. \cite{Telloni2022_Parker_Bepi, Palmerio2024}. Parker and Solar Orbiter in situ observations provide direct determination of particle velocity distributions, density, temperature, velocity, and magnetic field properties of the solar wind at some of the closest distances to the Sun, while Solar Orbiter also provides heavy ion composition diagnostics, to connect with properties and dynamics of the corona.

\subsubsection{Parker Solar Probe in situ: all solar wind properties}
With the launch of Parker in 2018, the heliophysics community is able to sample solar wind at distances previously inaccessible in situ. Parker has routinely observed solar wind thermal plasma (SWEAP; \citealt{Kasper2016_SWEAP}), magnetic and electric fields (FIELDS; \citealt{Bale2016}), and energetic particle (IS$\odot$IS; \citealt{McComas2016ISIS}) properties and their non-thermal characteristics below the orbit of Venus and within the Sun's Alfv\'en surface \citep{Kasper2021_subalfvenic}, the altitude where the local Alfv\'en speed is larger than the plasma's bulk speed. The probe reached its closest approach of its prime mission, 9.86 R$_{\odot}$, on December 24th, 2024 where it will continue to orbit at this perihelion distance \citep{Velli2020}. Parker measures the young solar wind, driving important model requirements to the early evolution of the solar wind. As indicated in Figure \ref{fig:FOVdensity_inner}, \ref{fig:FOVdensity_outer}, \ref{fig:FOVtemp_inner}, \ref{fig:FOVspeed_inner}, \ref{fig:FOVmagnetic_inner}, Parker's closest approach overlaps or nearly overlaps with the outer fields of view of several remote observations, verging on enabling truly direct remote-in situ connections. These figures also show Parker's in situ coverage has a relatively limited range in heliographic latitude (around $\pm4^{\circ}$). 

\subsubsection{Solar Orbiter in situ: all solar wind properties}
The Solar Orbiter mission provides concurrent in situ observations of the inner heliosphere with Parker. Solar Orbiter's perihelion reaches close to the orbit of Venus, $\sim$65R$_{\odot}$. Beginning in 2025, Solar Orbiter will climb farther out of the ecliptic in an inclined orbit around the Sun. 2025 March 31 will mark the first perihelion at a heliographic latitude of $\pm17^{\circ}$ and will later reach $\pm24^{\circ}$ and $\pm33^{\circ}$ in its extended mission phase (indicated in Figure \ref{fig:FOVdensity_outer}). As such, although not as close as Parker, Solar Orbiter's orbit and in situ data are highly complementary; it will sample a much larger range of heliographic latitudes and will probe solar wind from solar sources that would otherwise only be observed remotely. Its orbit will also create opportunities for more diverse spacecraft lineups with Parker and 1 au missions. As Parker, it measures particles \citep{Owen2020}, energetic particles \citep{Pacheco2020}, and fields \citep{Horbury2020} for direct diagnostics of the solar wind's density, temperature, speed, magnetic field, and non-thermal characteristics of particles. Additionally, it carries the Heavy Ion Sensor \citep{Livi2023} that measures heavy ion species of the solar wind that is important for connecting the Sun to the heliosphere, a brief review of heavy ion diagnostics can be found in \cite{Rivera2022}.

\subsection{Remote Coronal and Heliospheric Observations} \label{sec:remote}
%Further, obtaining 3D information typically requires the application of rotational tomography over a solar cycle, as discussed later.
There are several multi-wavelength coronal observations of the Sun from ground- and spaced-based instruments that can provide regular diagnostics of density, temperature, velocity, and magnetic field properties as well as non-thermal features of the full solar corona in connection with in situ observations. Quantifying properties such as non-thermal broadening of spectral lines, along with the plasma state and magnetic field, can inform on wave dynamics in the corona, important for coronal heating and wave energy flux calculations \citep{Hahn2014, Morton2019_alfvenflux}. The advantage of observations that include diagnostics of the full corona simultaneously, especially the ones listed here, is that: 1) they provide important solar context of source region conditions to probe solar wind formation, 2) in many cases where observations are coupled, they offer an continuous coronal view to investigate the solar wind's thermodynamic evolution across their FOV, and 3) provides a quantitative basis to directly compare the physical conditions, non-thermal line characteristics, and magnetic environment of the extended corona globally to help develop more realistic 3D models of the Sun. Instruments with full coronal coverage also provide more opportunities to connect remote coronal observations with in situ observations of the solar wind whose source regions can map to a large range of latitudes; this will be especially important as Solar Orbiter's orbit becomes more inclined. 

\subsubsection{Solar Orbiter remote observations: density, velocity}
Solar Orbiter, in addition to its in situ capabilities discussed above, carries a suite of imaging and spectrograph telescopes off the Sun-Earth line that provide direct solar coverage at some of the highest spatial resolutions ever achieved. In particular, Solar Orbiter provides extended coverage of the corona with images taken by the Metis coronagraph in H {\footnotesize I} Ly$\alpha$ and polarized Brightness (pB) probing the density and outflow speed of solar wind between 1.7-3.6R$_{\odot}$ at perihelion and 4.2---9R$_{\odot}$ near aphelion \citep{Romoli2021_Metis}. Although not included in the figures, the Full Sun Imager (FSI; \citealt{Rochus2020}) and High Resolution Imager (HRI) provide EUV coverage of the disk and its extended atmosphere important for coronal morphology. Additionally, Solar Orbiter also carries a UV spectrograph, Spectral Imaging of the Coronal Environment (SPICE; \citealt{SPICE2020}), that can scan a small FOV on the disk that covers a temperature range between $0.02-1$MK that contains several emission lines that can be utilized for some measurement of elemental composition at the Sun \citep{Brooks2022, Brooks2024_comp}. In particular, the elemental compositional signatures from SPICE can be important for linking the low corona to in situ observations from HIS in the heliosphere to confirm footpoint mapping predictions \citep[e.g. ][]{Yardley2024, Rivera2025}. 

\subsubsection{SOHO and STEREO coronagraphs: density}

As Metis, there are several other coronagraphs that can probe the density of the surrounding corona in broadband white light with observed total and pB, as shown in Figure \ref{fig:FOVdensity_inner} \citep{vandeHulst1950_pB}. COR 1 and COR 2 from the Sun Earth Connection Coronal and Heliospheric Investigation (SECCHI; \citealt{Howard2008_SECCHI}) suite on the STEREO A spacecraft regularly provide pB images of the extended region around the Sun, between 1.5-4 and 2-15R$_{\odot}$, respectively. In concert with observations from the Earth-Sun L1 point, the orbit of STEREO A also allows a multi-perspective view of the Sun in pB and coronal densities. Located at L1, the Large Angle and Spectrometric COronagraph (LASCO; \citealt{Brueckner1995LASCO}) C2 \& C3 are white light coronagraphs on Solar and Heliospheric Observatory mission (SOHO; \citealt{Domingo1995}) that cover 1.5--6R$_{\odot}$ and 3.7--30R$_{\odot}$, respectively. LASCO C2 also routinely measures pB for density diagnostics \citep{Romoli1997_pB}. 

Coronal tomography methods using density diagnostics from white light pB measurements have also extended our 2D view of the corona to 3D. Density reconstructions have been performed with observations from several instruments, including LASCO C2, STEREO COR 1, and, most recently, Metis  \citep{Frazin2002_tomography,Frazin2007pB_LASCOC2,Wang2017_COR1tom, Lloveras2019, Vasquez2024}. We note, tomographic 3D density and velocity reconstructions have also been produced using radio scintillation \citep[e.g.][]{Jackson2011,Tiburzi2023}, usually either using a coronal model or in situ data at 1au to normalize the inferred values to a physical range. This approach uses discrete radio point sources yielding sparse coverage but with interpolation can achieve global coverage and reach out in the heliosphere which with regular observations would be highly relevant in the context this paper.  

%This technique has also been used using instruments with non-global coverage but with a extended FOV e.g. the Solar Mass Ejection Imager \citep{Jackson2011} and potentially soon with the Wide Field Imager for Solar Probe (WISPR; \citealt{Vourlidas2016WISPR}) on Parker \citep{Kenny2024}. 

\subsubsection{Proba-3: density}
Additionally, although many coronagraphs have exclusion zones above the solar surface, there are several instruments that observe the solar atmosphere very close to the limb (below 2R$_{\odot}$) from space and the ground, as discussed further in Section \ref{sec:MLSO}. Proba-3, launched on December 5, 2024, is a mission that creates artificial total solar eclipses in space by performing precise formation flying of two satellites \citep{Shestov2021Proba3}. It will observe a white light pass band between 5350--5650 \AA~ and pB, providing density diagnostics in the low to middle corona, Figure \ref{fig:FOVdensity_inner}. It also observes two narrow passbands centered at Fe {\footnotesize XIV} 5304 \AA~ and another centered at  He {\footnotesize I} D3 5877 \AA~ between 1.1--3R$_{\odot}$. 

\subsubsection{MLSO: density, magnetic field, non-thermal broadening} \label{sec:MLSO}
A host of ground-based telescopes provide spectral observations of the Sun and its extended atmosphere to connect with orbiting spacecraft, here we highlight just a few in regular operation: The Mauna Loa Solar Observatory (MLSO) includes the Upgraded Coronal Multi-channel Polarimeter (UCoMP; \citealt{Landi2016}, upgraded version of the early generation CoMP; \citealt{Tomczyk2008CoMP}) and K-coronagraph (K-Cor; \citealt{deWijn2012_Kcor}). UCoMP observes several lines in the visible and near-infrared spanning 530--1083nm globally from 1.05 out to 1.95R$_{\odot}$. UCoMP provides intensity of several lines as well as linear polarization of Fe {\footnotesize XIII}. Polarimetric signatures in emission lines, such as in Fe {\footnotesize XIII}, are encoded with the strength and direction of the magnetic field in the solar atmosphere which can be used to probe the coronal magnetic field \citep{Gibson2017_nullpoints}. Also, coronal seismology has been effectively used to derive magnetic field properties using UCoMP observations \citep{Tomczyk2007_alfvenwaves,Yang2024}. Figure \ref{fig:FOVmagnetic_inner} highlights several operational and planned instruments for magnetic field diagnostics, including UCoMP. Additionally, UCoMP contains density-sensitive line pairs of Fe {\footnotesize XIII} to derive global coronal density, as shown in Figure \ref{fig:FOVdensity_inner}, which can also be used to derive non-thermal speeds, Figure \ref{fig:FOVmagnetic_inner}. MLSO's K-Cor is a coronagraph ranging between 1.05--3R$_{\odot}$ that provides observations of pB, important for density diagnostic of the corona, as included in Figure \ref{fig:FOVdensity_inner}. 

Although, containing a smaller FOV and not included in the figures, Cryo-NIRSP on the Daniel K. Inouye Solar Telescope (DKIST; \citealt{Rimmele2020}) can provide detailed, highly complementary observations to MLSO and Proba-3. DKIST is a 4-meter aperture off-axis Gregorian solar telescope at the summit of Haleakala on the island of Maui, Hawai’i. DKIST is composed of a five-instrument suite of telescopes to provide detailed coverage and spectropolarimetric observations of the solar surface and low corona in visible and near-infrared light. DKIST observes the Sun in unprecedented detail, aimed at deepening our understanding of the Sun's magnetic environment. Most recently, the Cryo-NIRSP telescope demonstrated the effect of Zeeman line splitting in off-limb spectropolarimetric observations that can be used to derive details of the magnitude and morphology of the coronal magnetic field \citep{Schad2024}. The wavelength range of Cryo-NIRSP also contains density-sensitive line pairs (Fe {\footnotesize XIII}) for density diagnostics and the spectral resolution to quantify non-thermal broadening in those lines. 

\subsubsection{Aditya-L1: magnetic field, non-thermal broadening}
The Aditya-L1 is an Indian mission, that launched in September 2023, located at Lagrange point L1, joining other missions like Advanced Composition Explorer (ACE; \citealt{Stone1998}), Wind \citep{Wilson2021}, SOHO, and Solar Dynamics Observatory (SDO) containing a set of both remote and in situ observations . Its payload contains several remote observations including spectrally pure coronal observations from the Visible Emission Line Coronagraph (VELC; \citealt{Mishra2024VELC_instru, Ramesh2024}), in Fe {\footnotesize XIV} 5303 \AA~(same as Proba-3), Fe {\footnotesize XI} 7892 \AA, Fe {\footnotesize XIII} 10747 \AA~ and white light continuum. In particular, VELC will capture polarimetric information of the Fe {\footnotesize XIII} 10747 \AA~ emission line that will inform on the coronal magnetic field, which is highly synergistic to UCoMP, DKIST, and future COSMO observations (see Section \ref{sec:future}), as shown in Figure \ref{fig:FOVmagnetic_inner}. Additionally, several of the lines provide information of non-thermal speeds \citep{Ramesh2024}.

\subsubsection{CODEX: density, temperature, velocity}
The recently deployed Coronal Diagnostic Experiment (CODEX; \citealt{Casti2024SPIE_CODEX, Gong2024SPIE_CODEX}) is an exciting new mission operating on the international space station (ISS). CODEX is composed of a coronagraph observing linearly polarized K-corona emission that covers a wavelength range of 385--440nm. The mission capitalizes on the coronal emission formed through Thomson scattered light to derive an electron density, temperature, and radial speed of the solar wind between 3-8R$_{\odot}$, included in Figures \ref{fig:FOVdensity_inner}, \ref{fig:FOVtemp_inner}, \ref{fig:FOVspeed_inner} \citep{Reginald2023}. CODEX provides highly complementary observations to continue to probe the solar wind state in tandem to observations at lower altitudes, e.g. UCoMP, Proba-3, K-Cor.

\subsection{Future and Planned Solar Missions}  \label{sec:future}

\subsubsection{PUNCH: density, velocity}
The Polarimeter to UNify the Corona and Heliosphere (PUNCH; \citealt{DeForest2022_PUNCH}), has a planned launch in spring of 2025. It contains a Narrow Field Imager (NFI) and a Wide Field Imagers (WFIs) that collectively covers an unprecedented FOV of 6-32R$_{\odot}$ and 20--180R$_{\odot}$, respectively. PUNCH will measure total brightness and pB, allowing for diagnostics of electron density and 3D location of the plasma in space to measure speed and propagation, as indicated in Figure \ref{fig:FOVdensity_inner}, \ref{fig:FOVdensity_outer}, \ref{fig:FOVspeed_inner}. PUNCH observations cover a large part of the corona and heliosphere that has historically been explored either remotely or in situ, making these observations incredibly important for linking coronal phenomena to the inner heliosphere.

\subsubsection{COSMO: density, temperature, non-thermal broadening}
The COronal Solar Magnetism Observatory (COSMO), from NSF/NCAR's High Altitude Observatory, is a planned/in development ground-based observatory that builds upon the current MLSO observatories. COSMO will include a larger version to present UCoMP with improved polarimetric capabilities, as well as the inclusion of a K-Cor white-light instrument, and a chromospheric instrument \citep{Tomczyk2016_COSMO}. Together, the COSMO observatory will be designed to routinely observe details of the magnetic field (including circular polarization as has been observed with DKIST), thermal and non-thermal spectroscopic properties to capture thermodynamics and wave dynamics of the global corona, as shown in Figure \ref{fig:FOVdensity_inner}, \ref{fig:FOVtemp_inner}, \ref{fig:FOVmagnetic_inner}.

\subsubsection{ECCCO: density, temperature}
A mission designed for full coronal diagnostics (including the solar disk) is the EUV CME and Coronal Connectivity Observatory (ECCCO; \citealt{Reeves2023_ECCCO}). ECCCO is a NASA Small Explorer mission currently in Phase A. It is composed of two instruments: an imager with a FOV out to 3R$_{\odot}$ and a spectrograph that produces spectrally pure, overlapped images of the full Sun. Both instruments cover a wavelength range of 171--205\AA~ sensitive to temperatures of 1--2.5MK, and 125--148\AA~ capturing hotter range of plasma at 8--12MK. The long wavelength channel contains density sensitive line pairs to probe density in the full corona while containing several consecutive Fe ions for robust temperature diagnostics of this region that is highly under observed in the EUV, as shown in Figure \ref{fig:FOVdensity_inner}, \ref{fig:FOVtemp_inner}.

\subsection{Solar Eclipses}
Although infrequent (and therefore not included in the figures), eclipses also provide uniquely powerful and detailed diagnostics of the corona despite the observations being limited in both duration and occurrence, taking place for a few minutes roughly once a year \citep{Habbal2021_eclipse}. Eclipses provide some of the most detailed spectroscopic coverage of the extended corona, as the moon fully occults the solar disk to block most of the light, revealing the faint corona. Ground-based observations can span out to 15–20 R$_{\odot}$ in white light, while spectroscopic imaging observations, e.g. Fe {\footnotesize XI} and Fe {\footnotesize XIV}, can be observed out to 3$R_{\odot}$. Despite their temporal limitations, eclipse observations have provided crucial information about coronal morphology, ion freeze-in distances, elemental composition, temperature, and density diagnostics \citep{Habbal2010temp, Habbal2011, Boe2018, DelZanna2023eclipse}. Therefore, should be incorporated into connection studies whenever possible \citep{Rivera2024AGU}\footnote{\url{https://www.cosmos.esa.int/web/solar-orbiter/-/science-nugget-coordinated-observations-during-the-2024-total-solar-eclipse}}.

\subsection{Present and Planned Solar Wind Diagnostics}

As indicated in the summary figures, there are several instruments that could be coupled together to probe density diagnostics between the low corona and in situ observations of the solar wind, as shown in Figure \ref{fig:FOVdensity_inner}, \ref{fig:FOVdensity_outer}. In contrast, temperature diagnostics are very limited, where there is only one active mission (CODEX, currently in commissioning) that provides electron temperature from 3-8R$_{\odot}$, missing low coronal observations. One thing to note about the CODEX mission is that its home on the ISS restricts the periods in which coronal observations can be taken and it only has a planned operational period of a few months. With the inclusion of CODEX and PUNCH, there will be several instruments that could provide velocity diagnostics as close to the Sun as 1.7R$_{\odot}$ (Metis during Solar Orbiter's perihelion) to couple with in situ observations, see Figure \ref{fig:FOVspeed_inner}. Conversely, as magnetic field diagnostics have been notoriously difficult and sparse in the corona, presently only UCoMP and, recently deployed, VELC on Aditya-L1 will provide routine polarimetric observations with full coronal coverage from the ground and space. As shown in the Figure \ref{fig:FOVmagnetic_inner}, UCoMP and VELC measurements are made in the low corona, well below 2R$_{\odot}$, and strictly on the Sun-Earth line leaving a large gap between remote and in situ magnetic field properties as well as non-thermal features.

\section{What is Missing for Coordinated Coverage of Solar Wind Formation and Heliospheric Evolution}

Although in principle all the observations discussed in the previous section could be connected, the synchronous availability of all components is exceedingly rare: Beyond operational schedules, one of the main reasons that makes alignment difficult is that the optimal configuration for tracking radial outflows happens when remote observations of the extended corona can be coupled (near or around the Sun-Earth line for ground+space-based instruments) while orbiting spacecraft intercept the solar wind whose source region is within some longitudinal range of the observed limb, i.e. in a quadrature configuration. 

Although challenging, the growing number of solar observatories provide more and more opportunities for such connections. There have been several examples of studies that have capitalized on special spacecraft configurations that effectively connect solar structures with their heliospheric counterparts \citep[]{Telloni2021, Adhikari2022, dePablos2022, Telloni2023, Baker2023, Yardley2024, Rivera2025}. \cite{Baker2023} connects outflows at the periphery of an AR, as indicated by the strong Doppler shifts from the spectral observations taken by Hinode/EIS. The AR outflows were mapped to a period of Alfv\'enic slow solar wind measured in situ with Solar Orbiter near perihelion. Similarly, \cite{Rivera2025} models the thermodynamic evolution of distinct Alfv\'enic and non-Alfv\'enic slow solar wind from its coronal source to several places in the heliosphere that includes the influence from a wave pressure gradient. However, these studies were carried out with solar wind diagnostics from a single height in the corona to a single, or sometimes several points in the heliosphere, missing much of the evolution in between. Also, for the case of \cite{Rivera2025}, the solar wind energy flux budget could only be assessed in the heliosphere because many of the diagnostics required to compute the kinetic, thermal, gravitational, and wave energy fluxes in the corona were not available.

At higher coronal altitudes, Metis has provided more extended coverage of the corona and solar wind within 9R$_{\odot}$ to map to the heliosphere. For example, \cite{Adhikari2022} connects observations of solar wind between 3.5–6.3R$_{\odot}$ using remote Metis observations that were linked with in situ observations from Parker ($\sim23.2$R$_{\odot}$) while in quadrature. This specific spacecraft alignment enabled an examination of the radial profile of a single slow solar wind stream that provided valuable constraints to a turbulence-driven solar wind model. However, although Metis provides diagnostics of plane of sky speed and density, no magnetic and temperature information below Parker's orbit was available, making it difficult to rigorously constraint those properties of the plasma at lower heights in the corona, and source region. 

An important aspect of studies that link solar wind to the corona, such as the ones listed, is connection science. A key measurement for connection science which has been under sampled in the inner heliosphere is the heavy ion composition of the plasma. In situ heavy ion properties such as the elemental composition are a powerful diagnostic for linking to sources of the solar wind as its relative abundances remain imprinted with the coronal source's heavy ion characteristics even after leaving the Sun \citep{vonsteiger2000, Ko2006, Brooks2015NatCo}. Apart from connection science, heavy ions measured in situ provide the detailed thermal structure of the corona, often used as a sensitive constraints to coronal heating and solar wind conditions in MHD models \citep{Oran2015, Lionello2019, Szente2022, Lionello2023, Riley2025}. As was mentioned in \ref{sec:solarmissions} and \ref{sec:remote}, the solar wind's source region can be more rigorously confirmed by matching the in situ and remotely derived elemental composition which is preserved from the corona to the heliosphere, together with other properties such as the magnetic polarity. However, these observations are not derivable beyond the solar limb and are limited to a small FOV with current and planned instruments.

\section{Summary and Final Remarks}
The paper summarizes the accessible and nearly available observations for density, temperature, velocity, magnetic field, and non-thermal diagnostics between the corona and heliosphere, Figures \ref{fig:FOVdensity_inner}-\ref{fig:FOVmagnetic_inner}. We note that integrating these measurements are limited to when the remote observations are taken in quadrature with in situ observations of the inner heliosphere as well as restricted to overlapping operation. 

Generally, there can be excellent coverage of density and velocity diagnostics between the low corona and the inner heliosphere when observations can be coupled. However, temperature, magnetic field 
and non-thermal diagnostics are much more limited to the low corona, below 2R$_{\odot}$, leaving a large observational gap between their in situ counterparts. Overall, more comprehensive coverage of ions and electron temperature, velocity, elemental composition, magnetic field, and non-thermal properties in the extended corona is needed. Off-limb spectroscopy, like in the era of SOHO's UltraViolet Coronagraph Spectrometer (UVCS; \citealt{Kohl1995UVCS}), for the full Sun would be ideal for merging with contemporary in situ observations made with Parker and missions such as ECCCO and COSMO. Additionally, since global observations of the coronal magnetic field have been limited to the low corona, moving towards making closer in situ measurements of the magnetic field would be highly valuable \citep{Krasnoselskikh2022, Viall2023_solardivers}.

Collectively, an international collaboration between the current solar fleet with the addition of upcoming new missions will provide more comprehensive coverage of the solar wind between remote and in situ observations, better bridging the solar-heliospheric system. Coupling the different missions can more effectively probe the thermodynamic and magnetic environment across the low to the middle and outer corona and beyond. Detailed and continuous coverage of the corona and solar wind is crucial to informing and advancing the current state of MHD models of the Sun and connection science capabilities that require time-dependent solutions, with self-consistent photospheric evolving boundary conditions, to capture more realistic coronal dynamics. In the advent of time-dependent global MHD modeling of the Sun \citep{Mason2023, Lionello2023}, it becomes critical to have higher time resolution 3D observations of the corona, which is currently limited to around the timescale of a solar rotation (by way of coronal tomography, \citealt{Jackson2011}) due to having mainly a single LOS of the corona. Therefore, apart from having more complete radial coverage, as outlined in this paper, additional perspectives of the corona are needed, including over the Sun's poles \citep{Hassler2023, Raouafi2023_firefly}. In turn, more refined models of the Sun can ultimately improve our space weather predictability and forecasting efforts. 

\section{Acknowledgements}
A special thank you to Benjamin Alterman, Daniel Seaton, Kathy Reeves, and Paul Bryans for fruitful discussions on the diagnostics of different instruments and science applications. Y.J.R. and S.T.B. are partially supported by the Parker Solar Probe project through the SAO/SWEAP subcontract 975569.  

\bibliographystyle{aasjournal}

\end{document}